\documentstyle[12pt]{article}
\begin{document}
\
\vskip 1truecm
\rightline{PUPTH-1472 (1994)}
\rightline{Bulletin Board: hep-ph/9406304}
\vskip 3truecm
\leftline{\bf TAU LEPTONS AND TOP QUARKS IN A `NON-LOCAL' BARYOGENESIS}
\leftline{\bf AT THE ELECTROWEAK PHASE TRANSITION}

\bigskip

\vspace{\baselineskip}
\begin{tabbing}
\hspace{1.0in} \= Tomislav Prokopec \\[\baselineskip]
\>Princeton University\\
\>Joseph Henry Laboratories\\
\>Physics Department, PO Box 708\\
\>Princeton, NJ 08544, USA\\
\end{tabbing}
\smallskip

\vspace{\baselineskip}

\leftline{\bf ABSTRACT
\footnote{\rm This lecture is based on a collaboration with  Michael Joyce and
Neil Turok. The lecture was delivered at the NATO Advanced Research
Workshop on `ELECTROWEAK PHYSICS AND THE EARLY UNIVERSE' held at Sintra,
Portugal from March 22 to 25, 1994.}}
\begin{quote}

Baryon asymmetry can be generated at the  first order
electroweak phase transition, provided there is a CP violation on the bubble
wall. In this report we discuss the role of  leptons and quarks in
the `non-local' baryogenesis mechanism.
\end{quote}

\bigskip

It is quite a remarkable fact that baryogenesis may have occurred at the
electroweak phase transition \cite{KuzminRubakovShap} with a
minimal extension of the Higgs sector,
in order to provide for the necessary CP violation
\cite{TurokZadrozny}.

CP violation in a  two complex Higgs doublet model is secured by a CP-odd
scalar field which changes in a definite manner  on the bubble wall,
as specified by the Higgs potential.
This field couples through the anomaly,
producing a bias in the baryon violating processes, and
therefore the possibility of baryon production
\cite{Turok}.

Cohen, Kaplan and Nelson \cite{ckn}
have proposed a non-local `charge transport' mechanism:
the {\it top\/} quark asymmetry generated by a CP-violating reflection off
the wall propagates  far into the
unbroken phase, and biases the baryon production.

In comparison to a `local' baryogenesis
occurring on the wall, there are two main advantages
to the `nonlocal' baryogenesis:

(a) the anomalous baryon number violating rate is unsupressed in the unbroken
phase, while it becomes exponentially suppressed on the wall,

(b) the extent over which baryogenesis takes place is much larger, it is
specified by the properties of the  diffusion tail in front of the wall.

In this work we will focus on discussing the role of various particle species
in a `non-local' baryogenesis, in particular we will compare the roles of
the {\it top} quark and $\tau$-lepton. Before we proceed, we review the main
steps involved in the calculation of the asymmetry.

In an analytic treatment, the problem of calculating the baryon number can
be broken up into: (a) modeling the reflection off the wall, which sources
(b) the diffusion equations,  thus specifying the profiles of particle
species in
front of the wall; (c) the effects of the hypercharge screening on propagation
of the asymmetry, and finally, (d) integration of the baryon rate equation.

(a) {\it Reflection.\/} Treatment of the reflection problem using the
perturbative Dirac equation is valid only in the thin wall limit:
$Lm\le 1$, which for the {\it top\/} quark gives a bound: $L\le 1/T$.
Perturbative calculations suggest $L\approx 20-40/T$. For $\tau$-leptons
this bound is marginally satisfied.
When the wall is much thicker than the mean free path of a particle, the
reflected asymmetry is exponentially suppressed due to the loss of coherence
\cite{GavelaHuet}. In the thin wall limit the perturbative Dirac
equation gives the reflected chiral particle asymmetry proportional to the
mass squared, wall velocity ($v_w$), $CP$-violation in the wall
($\Delta\theta_{CP}$), and inversely
proportional to the thickness of the wall ($m_H$):
\begin{equation}
J_L(0)  = -J_R(0)
 \approx { {v_w   m^2 m_H \Delta\theta_{CP} } \over {4\pi^2} }
\label{current}
\end{equation}
Note that this is a purely chiral current, which, after it thermalizes,
diffuses in front of the wall. To calculate  the
distribution of the diffusing particles, we model the source to the
diffusion equations by the chiral current in eq.~(\ref{current}).

(c) {\it Diffusion.\/} Propagation of the asymmetry in front of the wall is
specified by the diffusion equations or particle densities $n_i$ in the rest
frame of the wall ($x=z-v_wt$)
\begin{equation}
D_i n_i^{\prime\prime} +v_W n_i^{\prime} \Sigma_A \Gamma_A \Sigma_j \mu_j =
J_i^{\prime}(x)
\label{diffusioni}
\end{equation}
where $v_w$ denotes the velocity of the wall, $D_i$ diffusion constant of the
particle species $i$, $\Gamma_A$ is the rate for a  process $A$,
and $\Sigma_j \mu_j$ is the sum of the chemical potentials on the external
legs of the process $A$.
The extension of the source $\xi_i$ is defined as $\xi_i J_i^0=\int J_i$.

(d) {\it Baryon rate equation.} The rate of baryon production in the
unbroken phase is determined by the following rate equation:
\begin{equation}
\dot B=-{N_f\Gamma_s\over T}{\Sigma_i}\mu_i, \quad
\label{rate}
\end{equation}
where $\Gamma_s=\kappa \alpha_w^4 T$ is the weak sphaleron rate ($\kappa\in
[0.1,1]$), $N_f=6$ is the number of the fermionic flavors,  and
${\Sigma_i}\mu_i$  is the sum over the chemical potentials
for the left-handed particles.
\bigskip

Now we are ready to discuss  the role of leptons and quarks in the
`non-local' baryogenesis. Because of their much
larger tree-level mass, {\it top}  quarks and $\tau$-leptons are
reflected much more abundantly, and therefore require no further
consideration.

Let us first discuss  the propagation of leptons.
The relevant diffusion equations for the third family chiral leptons
($L_L$ and $L_R$) in the wall rest frame are static:
\begin{eqnarray}
D_L L_L'' + v_w L_L' - \Gamma_{LR} (a L_L -b L_R) &= J_L^{\prime}\\
D_R L_R'' + v_w L_R' + \Gamma_{LR} (a L_L -b L_R) &=J_R^{\prime}
\label{diffusionlepton}
\end{eqnarray}
In the above we also include the dominant
decay channel, which converts the left-handed leptons into the right-handed
ones {\it via\/} emission of a Higgs. For leptons:
$\Gamma_{LR}\approx 0.3\alpha_wy_\tau^2T$, where $y_\tau$ is the
$\tau$-lepton Yukawa coupling constant, $\alpha_w$ the $SU(2)$ coupling
constant, $a\approx 1/2$, $b\approx 1$. (Here we assume that the Higgs
particles
do not significantly affect the lepton distribution function,  as it was
argued in\cite{JoyceProkopecTurokNEW}.)

{}From eq.~(\ref{rate}), we infer that the relevant  quantity to
calculate, which is proportional to the baryon number, is the integral
of the left-handed lepton number in front of the wall, which can be obtained
in quite a simple manner.
After summing  and integrating twice eqs.~(\ref{diffusionlepton}) we get
\begin{equation}
\int_{0}^{+\infty} L_L={b\over{a+b}}
{\int_{-\infty}^{+\infty} (J_L+J_R)\over v_w}
\label{diffusionlepiii}
\end{equation}
To obtain this expression we used the fact that the dominant contribution
to the lepton integral comes from the diffusion tail in front of the
wall, in which $aL_L=bL_R$, and also that any lepton number vanish at
the spatial infinity.
The above integral gives the correct result for the integrated lepton number
for slow walls, for which  the diffusion tail extends far into the unbroken
phase: $v_w^2 << \Gamma_{LR} D_R <1$.

Next we  discuss the {\it top\/} quark case. We can write the analogous
equations for the left-  and right-handed baryons ($B_L$ and $B_R$), with
the decay term changed to
$\Gamma_{ss} (B_L -B_R)$, where $\Gamma_{ss}$ is the strong sphaleron rate.
Nevertheless, the simple trick
applied in solving the lepton equations does not work, because the diffusion
constants for left and right handed quarks are roughly the same.
This has  as a consequence: $J_{B_L}\approx -J_{B_R}$, so that
$B=B_L+B_R$ is not directly sourced by any current, and there is no diffusion
tail for baryons in front of the wall. The main
source for the baryon number comes from the strong sphaleron tail
$(D_q/2\Gamma_{ss})^{1/2}$ for $B_L-B_R$. The result is
\begin{equation}
\int_0^\infty (B_L-B_R)=
{1\over 4} {\int (J_{B_L}-J_{B_R})\over{\sqrt{2\Gamma_{ss}D_q}}}
\label{diffusionquarkii}
\end{equation}
The ratio of the baryon numbers for the {\it top\/} quark and
the $\tau$-lepton cases then reads:
\begin{equation}
{B_{top}\over B_\tau}={1\over 4}{a+b\over b}{v_w\over{\sqrt{2\Gamma_{ss}D_q}}}
{\int (J_{B_L}-J_{B_R})\over \int (J_{L_L}-J_{L_R})}\approx
{3\over 8}{v_w\over{\sqrt{2\Gamma_{ss}D_q}}}
{D_q\over D_{\tau_R}}\bigl({m_t(T)\over m_\tau(T)}\bigr)^2
\label{baryonratio}
\end{equation}
There are two reasons for the {\it top\/} quark suppression.
{\it Firstly\/}, there is no diffusion tail for {\it top\/}s; the strong
sphalerons cut it off in front of the wall, giving a suppression  factor
of approximately 10. {\it Secondly,\/} the extent of the source current
(before it thermalizes) is larger for the right handed $\tau$-leptons
at least by the ratio of the diffusion constants $D_{\tau_R}/D_q\sim 50$.
In the case of thicker walls, this suppression tends to be stronger than
linear, because
the {\it top\/}s have the tendency to be thermalized before they escape
the region of the wall, and thus be lost as a source for the diffusion
equations.

Naively, these two sources of suppression might be outweighed by the
large ratio of {\it top\/}-to-$\tau$ mass: in the standard model
$m_t^2/m_\tau^2\sim 10^4$, such that the {\it top\/} still contributes more
to the asymmetry than the $\tau$. However, in the two Higgs doublet theory,
quarks and leptons may couple to the different Higgs, which may have
different zero temperature {\it vev\/}s, giving the possibility that
around the phase transition the {\it top\/}-to-$\tau$ mass ratio
be much larger than at the zero temperature. In this case
the $\tau$-lepton produces more baryons.

And as a concluding remark, we announce the result of the recent work
\cite{JoyceProkopecTurokNEW} on thicker walls (comparable to the
perturbative result: $20-40/T$).
We consider the case  when the wall thickness is much larger than the
mean free path of particles, which justifies the usage of the fluid
equations approximation, but still thinner than the relevant
diffusion length $D/v_w$.  We find that, besides the reflection,
there is a {\it classical force\/} due to the  $CP$-violation on the bubble
wall acting differentially on particle species, causing a diffusion tail in
an axial particle number in front of the wall.
We find that the effect is proportional to the mass squared of the particles
in consideration and {\it independent\/} on the diffusion properties
of the species, singling out the {\it top\/} quark as the main
candidate for baryogenesis!

{\it Acknowledgements.\/}
I thank Michael Joyce, Neil Turok and Larry McLerran for useful  discussions.
The work of T.P. is partially	supported by
NSF contract PHY90-21984, and the David and Lucile Packard
Foundation.


\begin{thebibliography}{99}
 \bibitem{KuzminRubakovShap} {V. Kuzmin, V. Rubakov
and M. Shaposhnikov, Phys. Lett. {\bf 155B}, 36 (1985);
F. Klinkhamer and N. Manton, Phys. Rev. {\bf D30}, 2212 (1984).}

 \bibitem{TurokZadrozny}
{N. Turok and J. Zadrozny, Phys. Rev. Lett.
{\bf 65}, 2331 (1990); N. Turok and J. Zadrozny,
Nuc. Phys. {\bf B 358}, 471 (1991);
L. McLerran, M. Shaposhnikov, N. Turok and M. Voloshin,
 Phys. Lett. { \bf  256B}, 451 (1991).}

 \bibitem{Turok}
{N. Turok, in {\it Perspectives in Higgs Physics},
{\it ed.\/} by G. Kane, pub. World Scientific, p. 300 (1992);
A. Cohen, D. Kaplan and A. Nelson, Ann. Rev. Nucl. Part. Phys.
{\bf 43} 27-70 (1993).}

 \bibitem{ckn}
{A. Cohen, D.  Kaplan and A.
 Nelson, Nuc. Phys. {\bf B373}, 453  (1992)};
{A. Cohen, D. Kaplan and A.  Nelson, Phys. Lett. {\bf B294}  (1992) 57. }



 \bibitem{GavelaHuet}
{M. B. Gavela, P. Hern\'andez,
J. Orloff, and O. P\`ene, {\it Standard Model CP-violation and
Baryon Asymmetry\/}, preprint CERN 93/7081, LPTHE Orsay-93/48, HUTP-93/A036,
HD-THEP-93-45 (1993)};
{Patrick Huet and Eric Sather,
{\it Electroweak Baryogenesis and Standard Model CP-violation\/},
SLAC-PUB-6479, April, 1994.}

 \bibitem{JoyceProkopecTurokNEW}
{Michael  Joyce, Tomislav Prokopec, and
Neil Turok, {\it Classical Baryogenesis at the Electroweak Phase Transition\/},
PUPT preprint (1994).}

\end{thebibliography}
\end{document}